\documentclass[universe,article,accept,pdftex,oneauthor ]{Definitions/mdpi} 

\firstpage{1} 
\pubvolume{1}
\issuenum{1}
\articlenumber{0}
\pubyear{2022}
\copyrightyear{2022}
\externaleditor{Academic Editors: Dr. Eleonora Di Valentino, Prof. Dr. Leandros Perivolaropoulos, Dr. Jackson Levi Said 
}
\datereceived{1 November 2022} 
\dateaccepted{24 November 2022} 
\datepublished{} 
\hreflink{https://doi.org/} 

\usepackage{epsfig}
\usepackage{epstopdf}
\usepackage{amsfonts}
\usepackage{amssymb}
\makeatletter
\let\c@lofdepth\relax
\let\c@lotdepth\relax
\makeatother
\usepackage{subfigure}
\usepackage{tabularx}

\Title{DE Models with Combined $H_0 \cdot r_d $ from BAO and CMB Dataset and Friends}

\TitleCitation{DE Models with Combined $H_0 \cdot r_d $ from BAO and CMB Dataset and Friends}

\Author{{Denitsa} 
 Staicova \orcidA{0000-0001-6139-3125}}

\AuthorNames{Denitsa Staicova}

\AuthorCitation{Staicova, D.}

\address[1]{%
Institute for Nuclear Research and Nuclear Energy, Bulgarian Academy of Sciences, Sofia, 1784 
, Bulgaria; dstaicova@inrne.bas.bg\\}

\abstract{It has been theorized that dynamical dark energy (DDE) could be a possible solution to Hubble tension. To avoid degeneracy between Hubble parameter $H_0$ and sound horizon scale $r_d$, in this article, we use their multiplication as one parameter $c/\left(H_0 r_d\right)$, and we use it to infer cosmological parameters for 6 models---$\Lambda$CDM and 5 DDE parametrizations---the Chevallier--Polarski--Linder (CPL), the  Barboza--Alcaniz (BA), the  low correlation (LC), the  Jassal--Bagla--Padmanabhan (JBP) and  the  Feng--Shen--Li-Li models. We choose a dataset that treats this combination as one parameter, which includes the baryon acoustic oscillation (BAO) data $0.11 \le z \le 2.40$ and additional points from the cosmic microwave background (CMB) peaks ($z \simeq 1090$). To them, we add the marginalized Pantheon dataset and GRB dataset. We see that the tension is moved from $H_0$ and $r_d$ to $c/\left(H_0 r_d\right)$ and $\Omega_m$. There is only one model that satisfies the Planck 2018 constraints on both parameters, and this is LC with a huge error. The rest cannot fit into both constraints. $\Lambda$CDM is preferred, with respect to the statistical measures.
}

\keyword{cosmological tensions; dynamical dark energy; baryon acoustic oscillation; Pantheon dataset; gamma-ray bursts} 

\begin{document}

\section{Introduction}
The quest for understanding cosmological tensions  has driven research for years now. It seems that the tension between the direct measurements of the Hubble constant ($H_0$) from the late universe (\cite{Freedman:2000cf,Riess:1998cb,Perlmutter:1998np, Riess:2020fzl, Riess:2022mme}) and that from the early universe (i.e., from measuring the temperature and polarization anisotropies in the cosmic microwave background~\cite{Troxel:2017xyo,Aghanim:2018eyx,Ade:2015xua,Dainotti:2021pqg}) is only aggravated with the increase in the precision and knowledge of the systematics of the data and has reached 5 $\sigma$.  This discrepancy has spurred many works, trying to resolve whether the dark energy is a constant energy density or with a dynamical behavior, and if so, of what origin, leading to many different theories and possible explanations~\cite{Benisty:2021wxi,Capozziello:2011et,Bull:2015stt,DiValentino:2021izs,Yang:2021flj,Schoneberg:2019wmt,DiValentino:2017gzb,DiValentino:2020zio,Perivolaropoulos:2021jda,Lucca:2021dxo, Colgain:2022nlb, Colgain:2022rxy}.

There are many dark energy (DE) parametrizations~\cite{Wang:2018fng,Reyes:2021owe,Colgain:2021pmf, 2108.04188} that can be used in the search for deviations from the cosmological constant, $\Lambda$.
Some of them fall in the group of early dark energy models~\cite{Pettorino:2013ia, Poulin:2018cxd, Lin:2020jcb, Smith:2022hwi, Smith:2020rxx}, which modify physics of the early universe. Others modify the late-time universe physics, such as in the phantom dark energy~\cite{DiValentino:2020vnx, Haridasu:2020pms} models, and emergent dark energy~\cite{Li:2019yem,Yang:2020ope},~add interaction in the DE sector, as in the interacting dark energy~\cite{Kumar:2017dnp,DiValentino:2019ffd, Yang:2019uzo}, or~add exotic species or scalar fields~\cite{Gogoi:2020qif,Sakstein:2019fmf,Tian:2021omz,Nojiri:2021dze,Seto:2021xua}. For~a review on the taxonomy of DE models, see Refs.~\cite{Escamilla-Rivera:2021boq, Motta:2021hvl, Yang:2021eud}. Finally, there is the generalized emergent dark energy (GEDE)~\cite{Yang:2021eud} found to be able to compete with $\Lambda$CDM for some BAO datasets~\cite{Staicova:2021ntm}. In~this work, we take the so-called dynamical dark energy parametrizations, which allow for a non-constant DE contribution, regardless of the origin behind it. We take a number of models, namely the Chevallier--Polarski--Linder (CPL), the~ Barboza--Alcaniz (BA), the~low correlation (LC), the~ Jassal--Bagla--Padmanabhan (JBP) and  the  Feng--Shen--Li--Li (FSLLI) model, and we use statistical measures to judge their performance in fitting the~data.

An important part of the tensions debate revolves around the role of the sound horizon at drag epoch $r_d$. At~recombination, after~the onset of CMB at $z_*  \simeq  1090$, the~baryons escape the drag of photons at the drag epoch, $z_d  \simeq  1059$ (Planck 2018~\cite{Aghanim:2018eyx}). This sets the standard ruler for the baryon acoustic oscillations  (BAO)---the distance ($r_d$) at which the baryon--photon plasma waves oscillating in the hot  universe froze at $z=z_d$. The~sound horizon at drag epoch is given by
\begin{equation}
r_d = \int_{z_d}^{\infty} \frac{c_s(z)}{H(z)} dz
,\end{equation}
where $c_s \approx c \left(3 + 9\rho_b /(4\rho_\gamma) \right)^{-0.5}$ is the speed of sound in the baryon--photon fluid with the baryon $\rho_b(z)$ and the photon $\rho_\gamma(z)$ densities, respectively~\cite{Aubourg:2014yra,Arendse:2019itb}.

Many papers discuss the relation between the $H_0$ and the sound horizon scale $r_d$ for different models~\cite{Aylor:2018drw,Pogosian:2020ded,Aizpuru:2021vhd}. Any DE model claiming to resolve the $H_0$ tension should also be able to resolve the $r_d$ tension since they are strongly connected~\cite{Jedamzik:2020zmd,Aizpuru:2021vhd,delaMacorra:2021hoh}. In~other words, setting a prior on $r_d$ has a very strong effect on $H_0$ and vice~versa. In~this paper, to~avoid this problem, we combine the $H_0$ and $r_d$ into one parameter. We choose measurements that combine the $H_0 $ and $r_d$ from the BAO and the prior distance from the CMB peaks~\cite{Wang:2013mha, Mamon:2016wow, Grandon:2018uoe, Chen:2018dbv, daSilva:2018ehn, Zhai:2018vmm,DiValentino:2020hov, Nilsson:2021ute, Yao:2022kub} and we use them to infer the cosmological parameters for $\Lambda$CDM and 5 DDE models. To~the BAO+CMB dataset, we add the gamma-ray bursts (GRB) dataset and the Pantheon dataset with similarly marginalized dependence on $H_0$ (and $M_B$). We do this to expand the redshift considered by the models. In~a previous work~\cite{Staicova:2021ntm}, we used a similar approach in which we integrated $H_0 r_d$ in the $\chi^2$ of the model, while here, we use them as one single quantity without modifying the $\chi^2$. In~the marginalized version, we saw an interesting possibility for some DE model to fit the data better than $\Lambda$CDM. We continue this investigation with new models and a new approach in this~paper.

Historically, the~approach of using the combination $H_0 r_d$ is not new.  It has been used in~\cite{LHuillier:2016mtc} with BAO and SN data to find consistency with the Planck 2015 best-fit $\Lambda$CDM cosmology; Ref.~\cite{Shafieloo:2018gin} used the BAO data to fit the growth measurement, again finding consistency with the Planck 2015; Ref.~\cite{Arendse:2019hev} used the Cepheids and the Tip of the Red Branch measurements to calibrate BAO and SN measurements and find significant tension in both $H_0$ and $r_d$, despite testing the   $\Lambda CDM$ and DE models ($EDE$, $wCDM$, pEDE). The~implication is that  modifications of the physics after recombination fail to solve both tensions. The~overall conclusion is that the $H_0$ tension should not be considered separately from the $r_d$ measurement implied by it~\cite{Knox:2019rjx}. In~the current work, we choose a different approach. We repeat the analysis on $H_0 r_d$ used in earlier works, but we also take the ratio $r_*/r_d$ as an independent parameter. This means that we do not use the known analytical formulas for them, but~instead we use MCMC to infer them. This avoids using explicit prior knowledge on the baryon load of the universe. This way, we avoid both the degeneracy on $H_0 r_d$ from the BAO data, but~also we do not use as a hidden prior the Planck~measurements. 

The plan of the work is as follows: Section~\ref{sec:theory} formulates the relevant theory.  Section~\ref{sec:method} describes the method. Section~\ref{sec:res} shows the results, and Section~\ref{sec:sum} summarizes the~results.

\section{Theory}
\label{sec:theory}
A Friedmann--Lema\^itre--Robertson--Walker metric with the scale parameter\linebreak $a = 1/(1+z)$ is considered, where $z$ is the redshift. The~evolution of the universe for it is governed by the Friedmann equation, which connects the equation of the state for the $\Lambda$CDM~background:
\begin{equation}
    E(z)^2 = \Omega_{r} (1+z)^4 + \Omega_{m} (1+z)^3 + \Omega_{k} (1+z)^2 + \Omega_{DE}(z),
    \label{eq:hzlcdm}
\end{equation}

\noindent where in standard $\Lambda$CDM, $\Omega_{DE}(z)\to \Omega_\Lambda$, with~the expansion of the universe\linebreak $E(z)= H(z)/H_0$, where $H(z) := \dot{a}/a$ is the Hubble parameter at redshift $z$, and $H_0$ is the Hubble parameter today. $\Omega_{r}$, $\Omega_{m}$, $\Omega_{DE}$ and $\Omega_{k}$ are the fractional densities of radiation, matter, dark energy and the spatial curvature at redshift $z=0$. We take into account the radiation energy density as $\Omega_r = 1 - \Omega_m - \Omega_{\Lambda} - \Omega_{k}$. The~spatial curvature is expected to be zero for a flat universe, $\Omega_k=0$, and we set it to zero because we focus on DE~models.

We will consider a number of different DE models, all of which will feature a dark energy component depending on $z$. This can be done with a generalization of the Chevallier--Polarski--Linder (CPL) parametrization~\cite{Chevallier:2000qy,Linder:2005ne,Barger:2005sb}:
\begin{equation}
\Omega_{DE} \left(z\right) = \Omega_{\Lambda}  \exp\left[\int_0^{z} \frac{3(1+w(z')) dz'}{1+z'}\right]
\label{eq:ol}
\end{equation}
which allows for three possible models from which we will consider only the CPL:
\begin{equation}
w(z)=w_0 + w_a \frac{z}{z+1}
\end{equation}
and $\Lambda$CDM is recovered for $w_0=-1, w_a=0$. 

\textls[-25]{To this parametrization, we add another model~\cite{Barboza:2008rh, Escamilla-Rivera:2021boq}, which is the Barboza--Alcaniz (BA) model~with}

\begin{equation}
    w(z)=w_0+z\frac{1+z}{1+z^2}w_1
\end{equation}

This model is good for describing the whole universe history because~it does not diverge for $z \to -1$. It gives
\begin{equation}
    \Omega_{DE}=\Omega_{\Lambda}(1+z)^{3(1+w_0)}{(1+z^2)}^{\frac{3w_1}{2}}.
\end{equation}

Next, we use the low correlation model (LC) \cite{Wang:2008zh,  Escamilla-Rivera:2021boq} with
\begin{equation}
    w(z)=\frac{(-z+z_c)w_0+z(1+z_c)w_c}{(1+z)z_c}
\end{equation}
where $w_0=w(0)$ and $w_c=w(z_c)$ where $z_c$ is the redshift at which $w_0$ and $w_z$ are uncorrelated. The~effective entry into the EOS is
\begin{equation}
\Omega_{DE}=
\Omega_\Lambda(1+z)^{(3(1-2w_0+3wa))} e^{\frac{9(w_0-wa)z}{(1+z))}}
\end{equation}
where, here, are replaced $w_c$ with $w_a$ for consistency with the other~models.

The Jassal--Bagla--Padmanabhan (JBP) parametrization~\cite{Jassal:2004ej, Motta:2021hvl}
\begin{equation}
    w(z)=w_0+w_1\frac{z}{(1+z)^2}
\end{equation}
which gives
\begin{equation}
    \Omega_{DE}=\Omega_\Lambda (1+z)^{3(1+w_0)}e^{\frac{3w_1z^2}{2(1+z)^2}}
\end{equation}
with $w_0=w(z=0)$ and  $w_1 = (dw/dz)_{|(z=0)}$.

Finally, we will also test the Feng--Shen--Li--Li parametrization~\cite{Feng:2012gf, Motta:2021hvl} which is divergence-free for the entire history of the universe. It has two cases:
\begin{align}
&w(z)^+=w_0+w_1\frac{z}{1+z^2}\\
&w(z)^-=w_0+w_1\frac{z^2}{1+z^2}
\end{align}
with the final contribution to the EOS of each of them being, accordingly,
\begin{equation}
\Omega_{DE}^\pm =\Omega_\Lambda(1+z)^{3(1+w_0)}e^{\pm \frac{3w_1}{2}\arctan(z)}(1+z^2)^{\frac{3w_1}{4}}(1+z)^{\mp \frac{3}{2}w_1}
\end{equation}

In this work, the~plus case (i.e., $\Omega_{DE}^+$) is denoted as FSLLI, and the minus case (i.e., $\Omega_{DE}^-$) is denoted as~FSLLII.

The distance priors provide effective information of the CMB power spectrum in two aspects: the acoustic scale $l_\textrm{A}$ characterizes the CMB temperature power spectrum in the transverse direction, leading to the variation of the peak spacing, and~the ``shift parameter'' $R$ influences the CMB temperature spectrum along the line-of-sight direction, affecting the heights of the peaks. The~popular definitions of the distance priors are~\cite{Komatsu:2008hk}
\begin{equation}
\begin{split}
l_\textrm{A} =(1+z_*)\frac{\pi D_\textrm{A}(z_*)}{r_s(z_*)} ,\\
R\equiv(1+z_*)\frac{D_\textrm{A}(z_*) \sqrt{\Omega_m } H_0}{c},
\end{split}
\label{la:Rz}
\end{equation}
where $z_*$ is the redshift at the photon decoupling epoch with $z_*  \simeq  1089$ according to the $Planck$ 2018 results~\cite{Aghanim:2018eyx}. $r_*$ is the co-moving sound horizon at $z=z_*$. {Ref.} \cite{Chen:2018dbv}
 derives the distance priors in several different models using $Planck$ 2018 TT,TE,EE $+$ lowE which is the latest CMB data from the final full-mission Planck measurement~\cite{Aghanim:2018eyx}.  {We use the correlation matrices given in {Table~1
} in~\cite{Chen:2018dbv} to obtain the covariance matrices for $l_A$ and $R$  corresponding to each model.}

 {The angular diameter distance,  $D_\textrm{A}$,  needed for both the distance priors and the BAO points, is given by}
\begin{align}\label{}
D_\textrm{A}
=\frac{c}{(1+z) H_0 \sqrt{|\Omega_{k}|}  } \textrm{sinn}\left[|\Omega_{k}|^{1/2}\int_0^z \frac {dz'} {E(z')}\right]\ ,
\end{align}
where $\textrm{sinn}(x) \equiv \textrm{sin}(x)$, $x$, $\textrm{sinh}(x)$ for $\Omega_{k}<0$, $\Omega_{k}=0$, $\Omega_{k}>0$, respectively. We see that for the measured $D_A/r_d$, one can isolate the variable $b=c/(H_0 r_d)$. Below, we set $\Omega_k=0$, so this formula simplifies to
\begin{equation}
\frac{D_\textrm{A}}{r_d}= \frac{b}{(1+z)}\int_0^z \frac{dz'}{E(z')}
 \label{v_model}
\end{equation}

Finally, for~the SN and GRB datasets, we define the distance modulus $\mu(z)$, which is related to the luminosity distance ($d_L = D_A(1+z)^2$), through
\begin{equation}
        \mu_B (z) - M_B = 5 \log_{10} \left[ d_L(z)\right] + 25  \,,
\label{eq:dist_mod_def}
\end{equation}
where $d_L$ is measured in units of Mpc, and~$M_B$ is the absolute~magnitude.

\section{Methods}
\label{sec:method}
In this paper, we use three datasets, which we treat differently. For~the BAO dataset, the~definition of  $\chi^2$, which we minimize, is the standard one since we do not use the covariance matrix for it.
\begin{equation}
\begin{split}
\chi^2_{BAO} = \sum_{i} \frac{\left(\vec{v}_{obs} - \vec{v}_{model}\right)^2}{\sigma^2},
\end{split}
\end{equation}
where $\vec{v}_{obs}$ is a vector of the observed points (i.e., the values of $D_A/r_d$ at each $z$ in {Table}~\ref{tab:data}), $\vec{v}_{model}$ is the theoretical prediction of the model calculated with Equation~(\ref{v_model}) and $\sigma$ is the error of each~measurement.

Additionally, we use the SN and the GRB datasets to further constrain the models. For~them, we use the following marginalized over $H_0$ and $M_B$ formula, taken from~\cite{Staicova:2021ntm} so that we avoid setting priors on $H_0$ and $M_B$.

Following the approach used in (\cite{DiPietro:2002cz,Nesseris:2004wj,Perivolaropoulos:2004yr,Lazkoz:2005sp}),  the~integrated $\chi^2$ is
\begin{equation}
\tilde{\chi}^2_{SN,  GRB} = D-\frac{E^2}{F} + \ln\frac{F}{2\pi},
\end{equation}
{for
}
\begin{subequations}
\begin{equation}
D = \sum_i \left( \Delta\mu \, C^{-1}_{cov} \, \Delta\mu^T \right)^2,
\end{equation}
\begin{equation}
E = \sum_i \left( \Delta\mu \, C^{-1}_{cov} \, E \right),
\end{equation}
\begin{equation}
F = \sum_i  C^{-1}_{cov}  ,
\end{equation}
\end{subequations}
where  $\mu_{}^{i}$ is the observed luminosity, $\sigma_i$ is its error,~$d_L(z)$ is the luminosity distance, $\Delta\mu =\mu_{}^{i} - 5 \log_{10}\left[d_L(z_i)\right)$, $E$ is the unit matrix, and~$C^{-1}_{cov}$ is the inverse covariance matrix of the dataset. For~the GRB dataset, $C^{-1}\to 1/\sigma_i^2$ since there is no known covariance matrix for it.  For~the Pantheon dataset, the~total covariance is defined as $C_{cov}=D_{stat}+C_{sys}$, where $D_{stat}=\sigma_i^2$ comes from the measurement and $C_{sys}$ is provided separately~\cite{Deng:2018jrp}. 
Note, in~the so-defined marginalized $\chi^2$, the values of $M$ and $H_0$ do not change the marginalized $\tilde{\chi}^2_{SN}$.

The final $\chi^2$ is
$$\chi^2=\chi^2_{BAO}+\chi^2_{CMB}+\chi^2_{SN}+\chi^2_{GRB}.$$

\section{Datasets}
The dataset we are using is a collection of points from different BAO\linebreak observations~\cite{BOSS:2016goe,BOSS:2016wmc, BOSS:2016hvq,Blake:2012pj,Carvalho:2015ica,Seo:2012xy,Sridhar:2020czy,DES:2017rfo,Tamone:2020qrl,Zhu:2018edv,Hou:2020rse,Blomqvist:2019rah,duMasdesBourboux:2017mrl}, to~which we add the CMB distant prior~\cite{Chen:2018dbv} and the data from the binned Pantheon dataset, which contain $1048$ supernovae luminosity measurements in the redshift range $z\in (0.01,2.3)$ \cite{Pan-STARRS1:2017jku,Pan-STARRS1:2017jku} binned into 40 points. The~GRB dataset~\cite{Demianski:2016zxi} consists of 162 measurements in the range $z\in [0.03351,9.3]$. 

To estimate the possible correlations in the BAO dataset, we use the methodology in~\cite{Kazantzidis:2018rnb,Benisty:2020otr}. This method avoids the use of N-body mocks to find the covariance matrices due to systematic errors and replaces it with an evaluation of the effect of possible small correlation on the final result. We add to the covariance matrix for uncorrelated points $C_{ii} = \sigma_i^2$ symmetrically a number of randomly selected nondiagonal elements $C_{ij}$. Their magnitudes are set to $C_{ij} =0.5 \sigma_i \sigma_j$, where $\sigma_i \sigma_j$ are the published $1\sigma$ errors of the data points $i,j$. We introduce positive correlations in up to 6 pairs of randomly selected data points (more than $25\%$ of the data). {Figure
}~\ref{fig:checkCov} in {Appendix} {\ref{Appendix}} shows the corner plots with different randomized points for all the models we employ in this article. From~the plots, one can see that the effect from adding the correlations is below $10\%$ on average. {This indicates that we can consider the chosen set of BAO points for being effectively~uncorrelated.}

To run the inference, we use a Monte Carlo Markov Chain (MCMC) nested sampler to find the best fit. We use the open-source {package
} $Polychord$ \cite{Handley:2015fda} with the $GetDist$ package~\cite{Lewis:2019xzd} to present the~results. 

The prior is a uniform distribution for all the {quantities}:
 $\Omega_{m} \in [0, 1.]$, $\Omega_{\Lambda}\in[0, 1 - \Omega_{m}]$, $\Omega_r\in[0,1 - \Omega_{m} - \Omega_{\Lambda}]$, $c/ (H_0 r_d) \in [25, 35]$, $w_0 \in [-1.5, -0.5]$ and $w_a \in [-0.5,0.5]$. Since the distance prior is defined at the decoupling epoch ($z_*$) and the BAO---at drag epoch ($z_d$), we parametrize the difference between $r_s(z_*)$ and $r_s(z_d)$ as $rat= r_*/r_d$, where the prior for the ratio is $rat \in [0.9, 1.1]$.

\section{Results}
\label{sec:res}

Figure~\ref{fig:bwwa_BAO1} (as well as the figures in the Appendix \ref{Appendix})
show the final values obtained by running MCMC on the selected priors, the~numbers being in Table~\ref{results} in the {Appendix \ref{Appendix}}, where also the corner plots can be found. We see that the models differ seriously in their estimations for the physical quantities $c/(H_0 r_d), \Omega_m$ and $r_d/r_s$, probably due to the very wide prior imposed on $\Omega_m$. 

\begin{figure}[H]
\begin{tabularx}{\textwidth}{|c|c|}
\hline 
&\\[-1ex]
\includegraphics[width=0.468\textwidth]{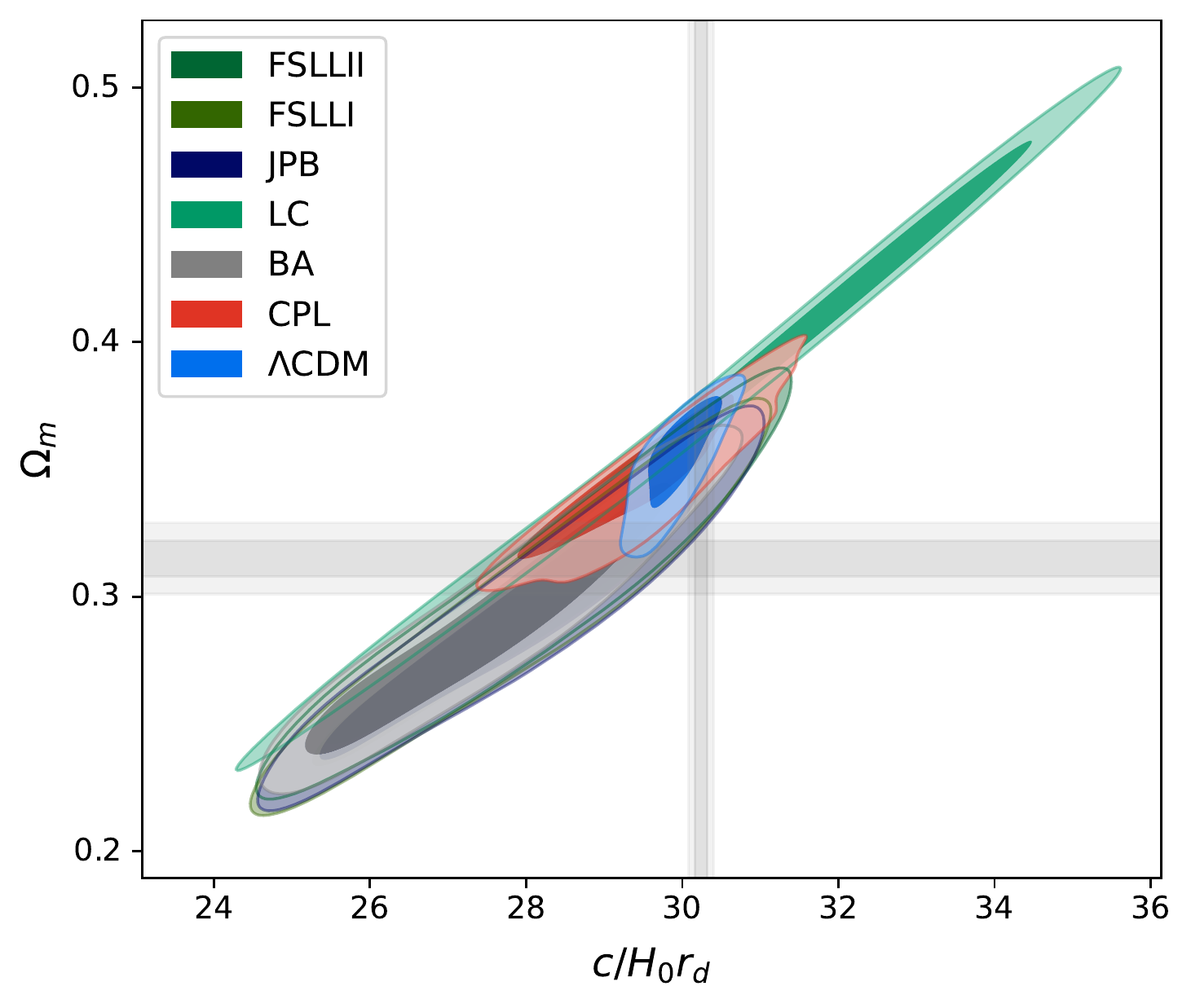}&
\includegraphics[width=0.468\textwidth]{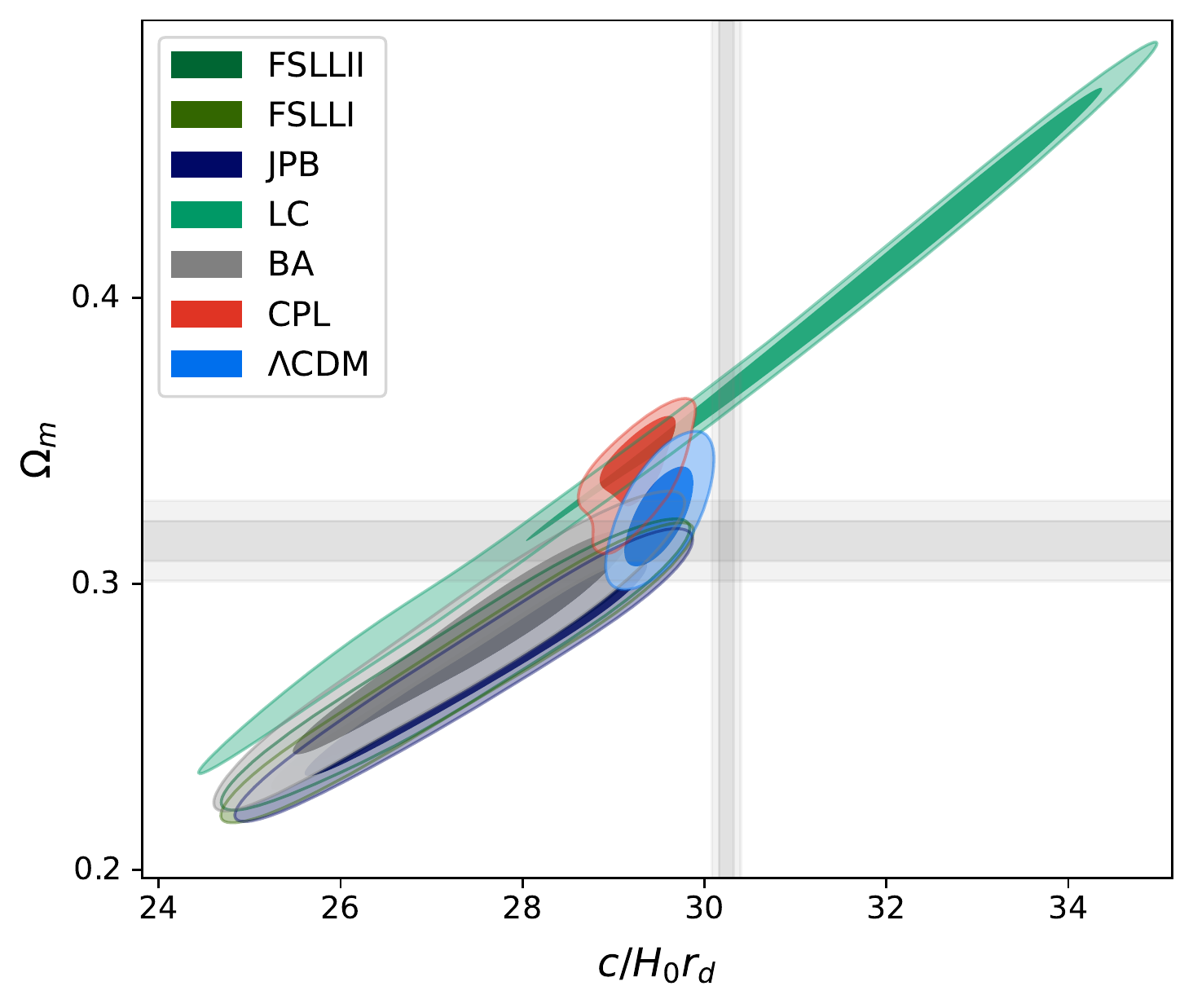}
\\
\includegraphics[width=0.468\textwidth]{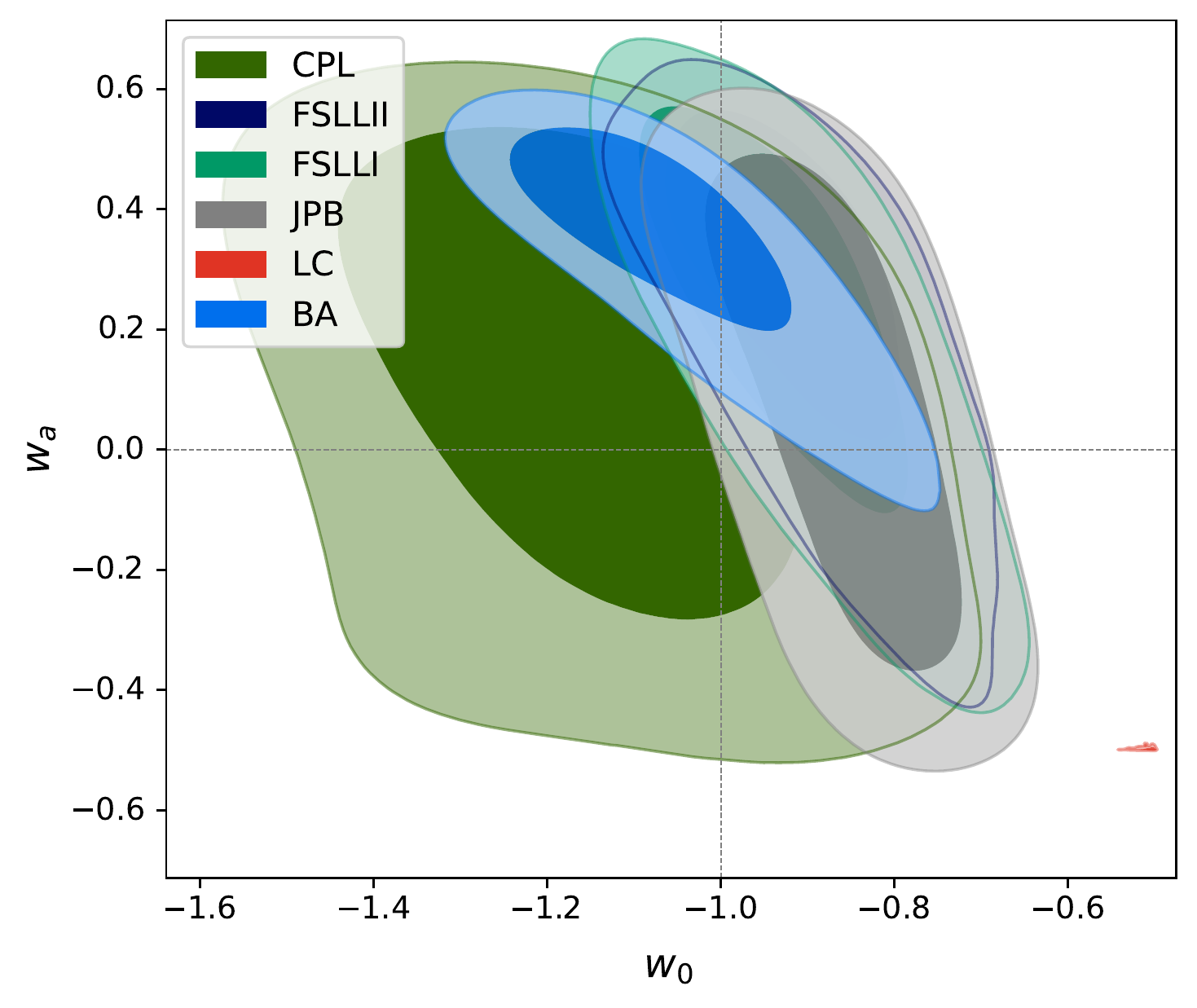}&
\includegraphics[width=0.468\textwidth]{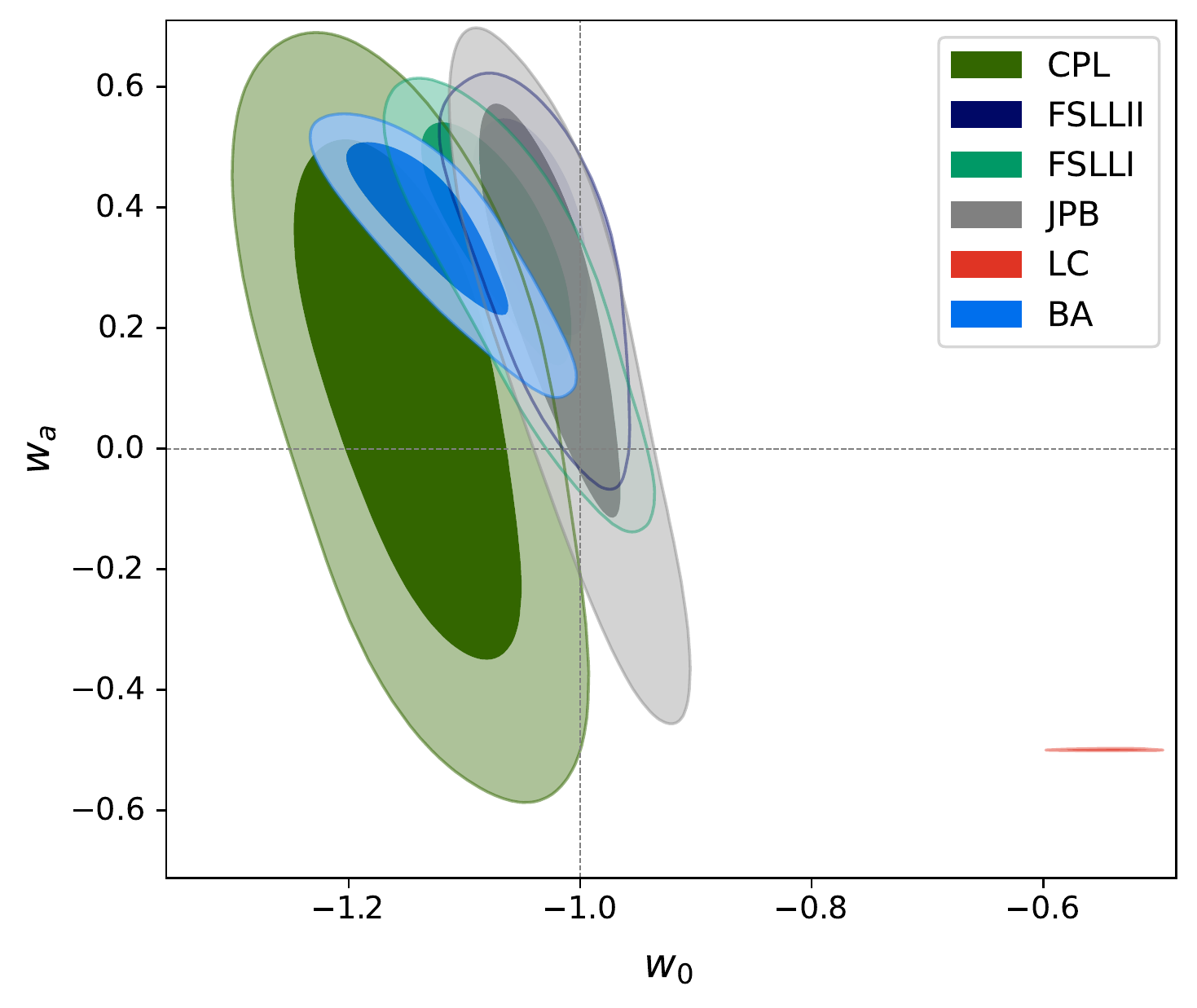}\\
\hline
\end{tabularx}
\caption{{The 2D contour plot for the different DE parametrizations for the BAO+CMB dataset to the left and for the BAO+CMB+SN+GRB to the right. The~upper panel shows the results for $\Omega_m$ vs. $c/(H_0 r_d)$ and the lower panel shows the results for $w$ and $w_a$. $\Lambda$CDM corresponds to $w_0 = -1$ and $w_a = 0$. The~grey lines show the $1 \sigma$ and $2\sigma$ of $\Omega_m$ and $c/(H_0 r_d)$ as measured by Planck 2018, while on the bottom plot the gray cross shows where we recover $\Lambda$CDM.}}
\label{fig:bwwa_BAO1}
\end{figure}

Since we avoid the degeneracy between $r_d$ and $H_0$ by considering the combined quantity $c/(H_0 r_d)$ this leads to an explicit correlation with $\Omega_m$ for some models and rather strict bounds on the error. The~values of $\Omega_m$ closest to the ones published by Planck 2018~\cite{Aghanim:2018eyx} $\Omega_m = 0.315\pm 0.007$ are for the BA, JPB and FSLLI models for the BAO dataset {and
} $\Lambda$CDM\endnote{In Sections~\ref{sec:res} and \ref{sec:sum} we discuss only the {\bf {flat} }$\Lambda$CDM model.
The~effect of the spatial curvature on DE models has been considered recently in~\cite{Yang:2022kho}.}  and BA for the BAO+SN+GRB. The~rest significantly overestimate $\Omega_m$ . For~the ratio $r_*/r_d$ Planck 2018 gives $0.98$, the~closest models are BA, JPB and FSLLI/FSLLII models for the BAO dataset and (flat) $\Lambda$CDM, JPB and FSLLI/FSLLII for the BAO+SN+GRB. For~$c/(H_0 r_d)$, the~Planck 2018 values is $30.26\pm0.06$. Here, the models closest to this value are $\Lambda$CDM, CPL and LC for the BAO dataset and $\Lambda$CDM, CPL and LC for the BAO+SN+GRB.

The DE parameters seem to be constrained to different level for the different models. As~a whole, the~trend to better constrain $w_0$ than $w_a$, which we observed in~\cite{Staicova:2021ntm} (and the referenced inside other works), is confirmed in this case as well. Notable exceptions are the BA and LC models, where the error of $w_a$ is much smaller. For them, however, the~other parameters seem to be outside of the expected boundaries.~$\Lambda$CDM performs as expected under both~datasets.

To compare the different models, we use well-known statistical measures. The~results can be seen in  Table~\ref{stats_BAO}. In~it, we publish four selection criteria:  Akaike information criterion (AIC), Bayesian information criterion (BIC), deviance information criterion (DIC) and the Bayes factor (BF). Since  for small datasets, both AIC and BIC are dominated by the number of parameters in the model (which are 3 for $\Lambda$CDM, and 5 for the DE models), we emphasize here on the DIC and the BF which rely on the numerically evaluated likelihood and evidence, making them more unbiased. The~DIC criterion, just like the AIC, selects the best model to be the one with the minimal value of the DIC measurement. The~reference table we use for DIC is
$\Delta DIC>10$ shows strong support for the model with lower DIC , $\Delta DIC = 5\text{--}10$ shows substantial support for the model with lower DIC, and $\Delta DIC<5$ gives ambiguous support for the model with lower DIC. Here, we use the logarithmic scale for the BF, for~which $ln(BF)>1$ shows support for the base model ($\Lambda$CDM), while $ln(BF)<-1$ for the other hypothesis. $|ln(BF)|<1$ shows an inconclusive~result.

\begin{table}[H] 
\caption{Selection criteria of different models in a comparison to the $\Lambda$CDM model for the BAO dataset and the BAO+SN+GRB {dataset
}.\label{stats_BAO}}
\newcolumntype{C}{>{\centering\arraybackslash}X}
\begin{tabularx}{\textwidth}{CCCCCCCC}
\toprule
\multicolumn{8}{c}{BAO+CMB}\\
			\midrule
			Model & AIC & $\Delta$AIC & BIC & $\Delta BIC$ & DIC & $\Delta$DIC & ln(BF) \\
			\midrule
			$\Lambda$CDM & 22.0 &  & 24.5 &  & 16.8 &  &  \\
			\midrule
			CPL & 25.7 & $-$3.7 & 29.9 & $-$5.4 & 16.5 & 0.3 & 0.6 \\
			\midrule
			BA & 25.3 & $-$3.3 & 29.5 & $-$4.9 & 16.2 & 0.65 & $-$5.3 \\
			\midrule
			LC & 56.0 & $-$33.9 & 60.2 & $-$35.6 & 51.1 & $-$34.3 & 38.5 \\
			\midrule
			JPB & 27.8 & $-$5.8 & 31.9 & $-$7.4 & 18.6 & $-$1.8 & $-$3.5 \\
			\midrule
			FSLLI & 27.1 & $-$5.1 & 31.3 & $-$6.9 & 17.9 & $-$1.1 & $-$3.8 \\
			\midrule
			FSLLII & 26.6 & $-$4.6 & 30.8 & $-$6.3 & 17.4 & $-$0.65 & $-$4.0\\
			\midrule
			\multicolumn{8}{c}{BAO+CMB+SN+GRB}\\
			\midrule
			$\Lambda$CDM & 228.1 &  & 238.3 &  & 222.7 &  &  \\
			\midrule
			CPL & 229.2 & $-$1.1 & 246.1 & $-$7.8 & 219.9 & 2.8 & $-$1.2 \\
			\midrule
			BA & 229.0 & $-$0.9 & 246.0 & $-$7.8 & 219.8 & 2.9 & $-$5.9 \\
			\midrule
			LC & 436.8 & $-$208.7 & 453.7 & $-$215.5 & 427.6 & $-$204.9 & 208.9 \\
			\midrule
			JPB & 232.2 & $-$4.1 & 249.2 & $-$10.9 & 222.9 & $-$0.2 & $-$4.0 \\
			\midrule
			FSLLI & 231.1 & $-$2.9 & 248.0 & $-$9.7 & 221.9 & 0.9 & $-$3.7 \\
			\midrule
			FSLLII & 230.5 & $-$2.4 & 247.4 & $-$9.2 & 221.3 & 1.5 & $-$4.8 \\
			\bottomrule
\end{tabularx}
\end{table}


From Table~\ref{stats_BAO}, we see that the AIC and BIC for all models show a preference for $\Lambda$CDM. For~the DIC criterion, we see a slight possibility for a preference for other models in the case of the CPL and BA models for both tested datasets. For~the BF, we see that there is some possible preference for BA, JPN and FSLLI/FSLLII for the BAO+CMB case and for CPL and BA, JPN and FSLLI/FSLLII in the BAO+CMB+SN+GRB case. The~results of the LC model show that it is underfitting the data (from the $\chi^2/dof$$\sim$$2$) and the statistics for it is not reliable. This demonstrates another benefit of performing the statistical~analysis.

The preference for the BA and LC models which we observe was also observed in the results of~\cite{Escamilla-Rivera:2021boq}, where the authors studied a dataset consisting of  SN, cosmic chronometers and gravitational~waves.

\begin{figure}[H]
\includegraphics[width=0.495\textwidth]{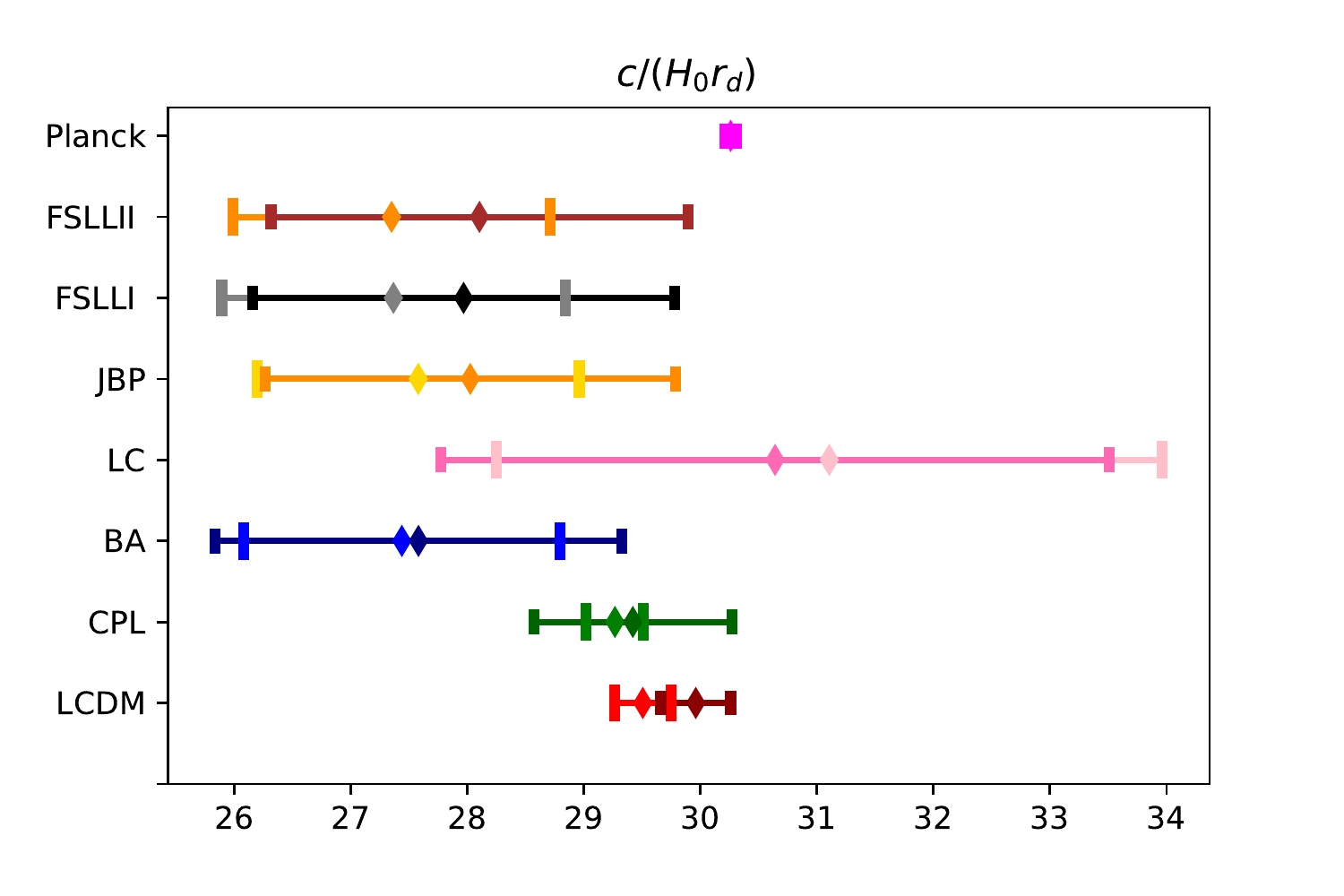}
\includegraphics[width=0.495\textwidth]{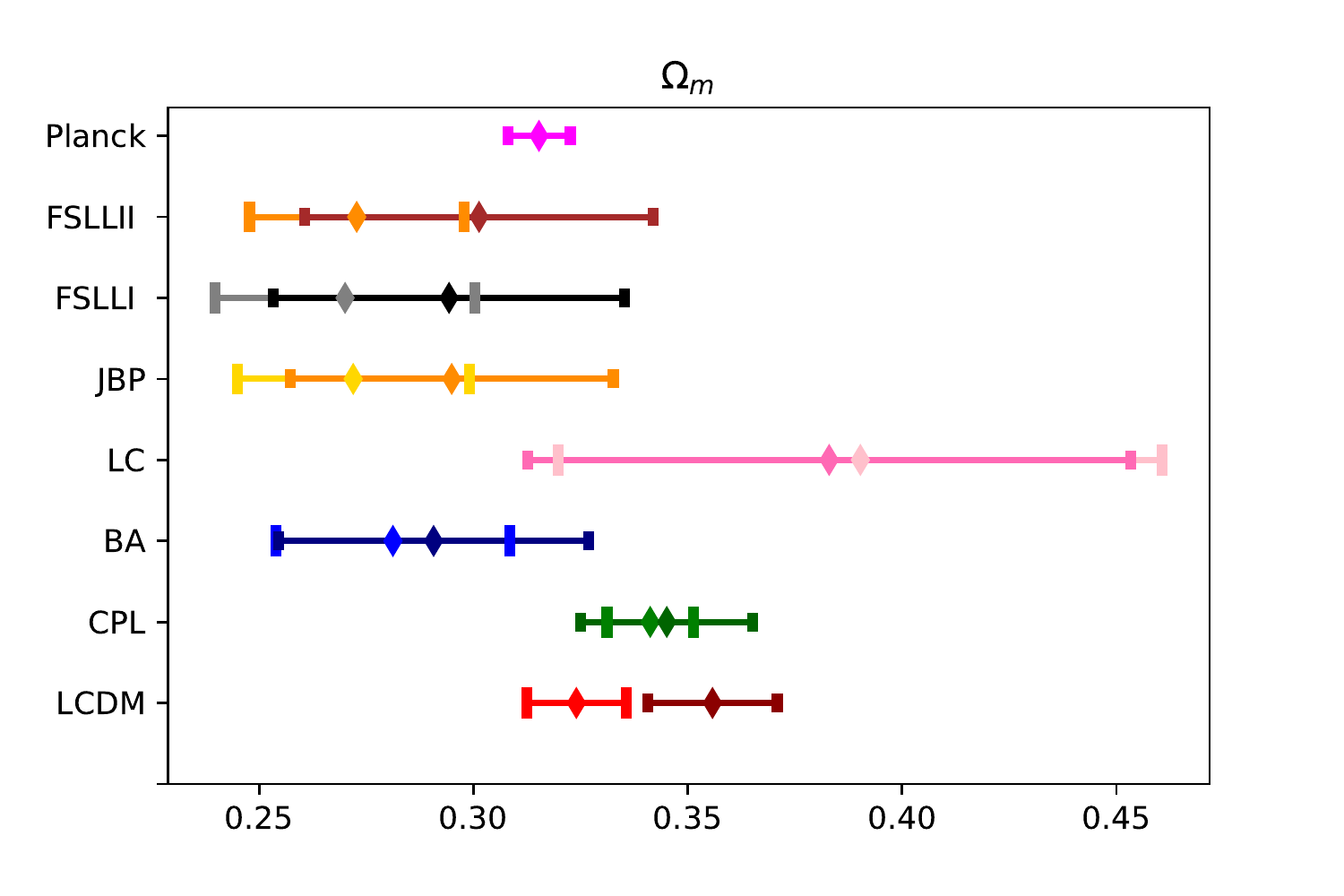}

\caption{{ {The final values of the $c/(H_0 r_d)$ from different DE models, compared to the values from Planck for the BAO+CMB+SN+GRB dataset. The~smaller, darker, errorbox are for BAO+CMB, the~lighter, bigger errorbox--for SN+GRB.}}}
 	\label{fig:H0rdvalGRB}
\end{figure}

The BAO dataset we use combines the $H_0$ and the $r_d$ into one quantity. Therefore, we estimate the new variable $c/(H_0 r_d)$$\sim$$30$. Figure~\ref{fig:H0rdvalGRB} shows the values of the $c/(H_0 r_d)$ for different models vs. the result from Planck 2018: $30.24 \pm 0.08$. For~comparison, the~most recent local measurement by SH0ES is $30.19 \pm 0.53$, corresponding to $H_0=73.01 \pm 0.99 \; \text{km s}^{-1} \text{Mpc}^{-1}$ \cite{Riess:2022mme}. We do not put it on the plot, because~the $r_d$ used to obtain it is the indirect result from inference on the H0LiCOW+SN+BAO+SH0ES dataset~\cite{Arendse:2019hev}. It is, however, clearly very close to the Planck value, as~expected.

On Figure~\ref{fig:H0rdvalGRB}, we superimpose the BAO+CMB-only result with the BAO+CMB+SN+GRB one. This figure enables us to visually track the tension between the Planck 2018 results and the datasets we use, which are mostly local universe ones (except for the 2 CMB points). We see that the tension is now between $c/(H_0 r_d)$ and $\Omega_m$. The~models whose bounds cross with the Planck 2018 one for $c/(H_0 r_d)$ are $\Lambda$CDM, CPL and LC for BAO+CMB and only LC for the BAO+CMB+SN+GRB dataset. For~$\Omega_m$, the models that enter the interval are all but $\Lambda$CDM and $CPL$ for the BAO+CMB dataset and  $\Lambda$CDM, BA, LC for the BAO+CMB+SN+GRB dataset. We see that the inclusion of the new datasets decreases the number of models satisfying the constraints. The~only model that is not in tension is LC because of its huge error. Notably, in~this approach, $\Lambda$CDM, while satisfying the bounds for $c/(H_0 r_d)$, does not satisfy them for $\Omega_m$. 

\vspace{-6pt}

From the plot, one can see that in general adding the new datasets decrease the errors, but~they do not move the mean values in the same direction and the overall effect is not very big. This may be due to unknown errors in the SN+GRB dataset or to the fact that this dataset is not sensitive toward the combined variable $c/(H_0 r_d)$ since we have marginalized over $H_0$ so that we do not have to impose a prior on $r_d$. Because~of this, the~only effect the SN+GRB dataset has on the combined variable is indirect, through $\Omega_m$ and the other parameters. It could also point to some inconsistency in the $\mu(M_B)$ relation such as the ones considered in~\cite{Benisty:2022psx,Ferramacho:2008ap,Linden:2009vh,Tutusaus:2017ibk,DiValentino:2020evt,Perivolaropoulos:2022khd}) questioning the assumption that $M_B=const$.

\section{Discussion}
\label{sec:sum}
This paper uses the combination $H_0\cdot r_d$ to avoid the degeneracy between $H_0$ and $r_d$ which has plagued the use of BAO measurements and could be part of the resolution of cosmological tensions. The~use of a combined parameter avoids imposing separate priors on $H_0$ and $r_d$ and thus it avoids additional assumptions on them. We use points from the late universe, the~BAO dataset ($z<2.4$)), few points from the early universe (the CMB distant priors, ($z  \simeq  1089$)), to~which we add SN data and GRB datasets, properly marginalized, to~make a statistical comparison between different DE~models.

The results show that the tension is now between the new parameter $c/(H_0 r_d)$ and $\Omega_m$---the only model that fits in the constraints set by Planck 2018 is LC, which comes with the biggest error. For~the rest of the models, one of the two parameters do not fit the constraints, even if some of them somewhat reduce the tension. Statistically, there is a preference for the $\Lambda$CDM model over the DE models in most cases. It is worth noting that there is strong evidence in support of $\Lambda$CDM compared to all other models only when using AIC and BIC, while from DIC and BF, the~support is not substantial, and it even slightly favors other models. This result raises the question of the use of different statistical measures when comparing DE models, and~also it opens the possibility that a better DE model may eventually help in reducing both the $H_0$ tension and the $r_d$ tension. 

Another interesting point is that for some models, the~known impossibility to constrain $w_a$ is eliminated and $w_a$ has very tight bounds. These models, LC and BA and somewhat FSLLI, show interesting new possibilities for DE models. Furthermore, the~choice of datasets and models make explicit the degeneracy between $H_0\cdot r_d$ and $\Omega_m$, emphasizing the need to find a way to disentangle the three quantities---$H_0, r_d$ and $\Omega_m$---if we are to understand the cosmological tensions. The~results show that adding the SN and GRB datasets decrease the errors on the constrained parameters, but~they do not move them in the same direction for each model. We see that combining different datasets and different marginalization techniques, along with the use of statistical measures, is a promising tool to study new cosmological~models.

\vspace{6pt}

\funding{{Bulgarian National Science Fund research grants KP-06-N58/5/19.11.2021.}
}

\dataavailability{All the data we used in this paper were taken from the corresponding citations and available to~use.} 

\acknowledgments{D.S. thanks David Benisty for the useful comments and discussions. D.S. is thankful to Bulgarian National Science Fund for support via research grants KP-06-N58/5.}

\conflictsofinterest{{The authors declare no conflict of interest.}

}




\appendixtitles{yes} 
\appendixstart
\appendix
\section[\appendixname~\thesection]{Some Extra~{Material}\label{Appendix}}


\begin{table}[H] 
\caption{{The} uncorrelated dataset used in this paper. For~each redshift, the table presents the parameter, the~mean value, and~the corresponding error bar. The~reference and the collaboration are also reported.\label{tab:data}}

\begin{adjustwidth}{-\extralength}{0cm}

\newcolumntype{C}{>{\centering\arraybackslash}X}
\begin{tabularx}{\fulllength}{CCCCm{8cm}<{\centering}C}
\toprule
$\mathbold{z}$   & $\mathbold{D_A/{r_d}}$ & \textbf{Error} & \textbf{Year}  & \textbf{Survey} &  \textbf{Ref.} \\
\midrule
$0.11$  & $2.607$ & $0.138$&  $2021$ & SDSS blue galaxies & \cite{deCarvalho:2021azj}\\
$0.24$   & $5.594$& $0.305$&  $2016$ &BOSS-DR12 RSD of LOWZ and CMASS & \cite{BOSS:2016goe}\\
$0.32$    &  $6.636$  &  $0.11$   &  $2016$  &  SDSS-DR9+DR10+DR11+DR12 +covariance  & \cite{BOSS:2016wmc} \\
$0.38$    &  $7.389$  &  $0.122$  &  $2019$  &  BOSS-DR12 power spectrum & \cite{BOSS:2016hvq}\\
$0.44$    &  $8.19$  &  $0.77 $  &   $2012$  &  WiggleZ (galaxy clustering)  & \cite{Blake:2012pj}\\
$0.54$    &  $9.212$  &  $0.41$    &  $2012$  &  SDSS-III DR8 (luminous galaxies) & \cite{Seo:2012xy}\\
$0.6$    &  $9.37$  &  $0.65$   &  $2012$  &   WiggleZ (galaxy clustering) & \cite{Blake:2012pj}\\
$0.697$    &  $10.18$  &  $0.52$   &  $2020$  &  DECals DR8 (LRG)  & \cite{Sridhar:2020czy}\\
$0.73$    &  $10.42$  &  $0.73$  &   $2012$  &  Wiggle (galaxy clustering) & \cite{Blake:2012pj}\\
$0.81$    &  $10.75$  &  $0.43$   &  $2017$  &  DES Year1  (galaxy clustering) & \cite{DES:2017rfo}\\
$0.85$    &  $10.76$  &  $  0.54$   &  $ 2020$  &  eBOSS DR16 ELG & \cite{Tamone:2020qrl}\\
$0.874$    &  $11.41$  &  $0.74$    &  $2020$  &  DECals DR8 (LRG) & \cite{Sridhar:2020czy}\\
$1.00$    &  $11.521$  &  $1.032$    &  $2019$  & eBOSS DR14 quasar clustering & \cite{Zhu:2018edv}\\
$2.00$    &  $12.011$  &  $0.562$  &   $2019$  &  eBOSS DR14 quasars clustering & \cite{Zhu:2018edv}\\
$2.35$    &  $10.83$  &  $0.54$  &  $2019$  &  BOSS DR14 Lya and quasars & \cite{Blomqvist:2019rah}  \\
$2.4$    &  $10.5$  &  $0.34$  & 2017 & SDSS-III/DR12 & \cite{duMasdesBourboux:2017mrl}\\
\bottomrule
\end{tabularx}

\end{adjustwidth}
\end{table}

\vspace{-10pt}
\begin{table}[H] 
\caption{{The} 68 $\% $ C.L. limits for $R$, $l_A$, in~different cosmological models and their correlation matrix for from Planck $2018$ $TT,TE,EE+lowE$; see the text for details.\label{distancePr}}
\newcolumntype{C}{>{\centering\arraybackslash}X}
\begin{tabularx}{\textwidth}{Cm{5cm}<{\centering}CC}
\toprule
$\Lambda$CDM & $Planck~ \textrm{TT,TE,EE}+\textrm{lowE} $ & $R$ & $l_\textrm{A}$ \\
\midrule
$R$  & $1.7502\pm0.0046$ & $1.0$&    $0.46$ \\
$l_\textrm{A}$  & $301.471^{+0.089}_{-0.090}$ & $0.46$&    $1.0$  \\
\midrule
  $w$CDM & $Planck~ \textrm{TT,TE,EE}+\textrm{lowE} $ & $R$ & $l_\textrm{A}$  \\
\midrule
$R$ & $1.7493^{+0.0046}_{-0.0047}$  & $1.0$&    $0.47$\\ \specialrule{0em}{1pt}{1pt}
$l_\textrm{A}$ & $301.462^{+0.089}_{-0.090}$ & $0.47$&    $1.0$ \\
\midrule
   $\Omega_k \Lambda$CDM & $Planck~ \textrm{TT,TE,EE}+\textrm{lowE} $ & $R$ & $l_\textrm{A}$\\
\midrule
$R$ & $1.7429\pm0.0051$ & $1.0$ & $0.54$  \\
$l_\textrm{A}$ & $301.409\pm0.091$ & $0.54$&   $1.0$\\
\bottomrule
\end{tabularx}
\end{table}

%

\begin{table}[H] 
\caption{{The} posterior values for $c/(H_0 r_d)$, $\Omega_m$, $r_*/r_d$ and $w_0,w_a$ for different parametrization of DE for the BAO+CMB dataset (top) and for the BAO+CMB+SN+GRB (bottom).\label{results}}
 \resizebox{\textwidth}{!}{  
    \begin{tabular}{cccccc}
\toprule
\multicolumn{6}{c}{BAO+CMB}\\
			\midrule
			Model & $c/(H_0 r_d)$ & $\Omega_m$ & $r_*/r_d$ & w & $w_a$ \\
			\midrule
			$\Lambda$CDM & $29.96\pm 0.3$ & $0.36\pm 0.02$ & $0.92\pm 0.01$ & $-$1.000 & 0.000 \\
			\midrule
			CPL & $29.42\pm 0.85$ & $0.35\pm 0.02$ & $0.91\pm 0.01$ & $-1.14\pm 0.19$ & $0.13\pm 0.31$ \\
			\midrule
			BA & $27.58\pm 1.74$ & $0.29\pm 0.04$ & $0.93\pm 0.02$ & $-1.06\pm 0.11$ & $0.35\pm 0.12$ \\
			\midrule
			LC & $30.64\pm 2.87$ & $0.38\pm 0.07$ & $0.901\pm 0.0009$ & $-0.5082\pm 0.0072$ & $-0.4979\pm 0.0018$ \\
			\midrule
			JPB & $28.03\pm 1.76$ & $0.29\pm 0.04$ & $0.94\pm 0.02$ & $-0.86\pm 0.09$ & $0.08\pm 0.31$ \\
			\midrule
			FSLLI & $27.97\pm 1.81$ & $0.29\pm 0.04$ & $0.94\pm 0.02$ & $-0.92\pm 0.1$ & $0.22\pm 0.24$ \\
			\midrule
			FSLLII & $28.1\pm 1.79$ & $0.3\pm 0.04$ & $0.94\pm 0.02$ & $-0.91\pm 0.09$ & $0.26\pm 0.2$ \\
			\midrule
		    \multicolumn{6}{c}{BAO+CMB+SN+GRB}\\
			\midrule
			$\Lambda$CDM & $29.51\pm 0.24$ & $0.32\pm 0.01$ & $0.95\pm 0.01$ & $-$1.000 & 0.000 \\
			\midrule
			CPL & $29.27\pm 0.25$ & $0.34\pm 0.01$ & $0.91\pm 0.01$ & $-1.15\pm 0.06$ & $0.09\pm 0.31$ \\
			\midrule
			BA & $27.44\pm 1.36$ & $0.28\pm 0.03$ & $0.94\pm 0.01$ & $-1.13\pm 0.04$ & $0.37\pm 0.1$ \\
			\midrule
			LC & $31.11\pm 2.86$ & $0.39\pm 0.07$ & $0.9009\pm 0.0007$ & $-0.55\pm 0.02$ & $-0.4992\pm 0.0007$ \\
			\midrule
			JPB & $27.58\pm 1.38$ & $0.27\pm 0.03$ & $0.9657\pm 0.0084$ & $-1.02\pm 0.04$ & $0.22\pm 0.23$ \\
			\midrule
			FSLLI & $27.37\pm 1.47$ & $0.27\pm 0.03$ & $0.9614\pm 0.0082$ & $-1.06\pm 0.04$ & $0.32\pm 0.14$ \\
			\midrule
			FSLLII & $27.35\pm 1.36$ & $0.27\pm 0.03$ & $0.9556\pm 0.0099$ & $-1.04\pm 0.03$ & $0.35\pm 0.13$ \\
			\bottomrule
 \end{tabular}%
    }
\end{table}

\begin{figure}[H]
 	
(\textbf{a}) \includegraphics[width=0.25\textwidth]{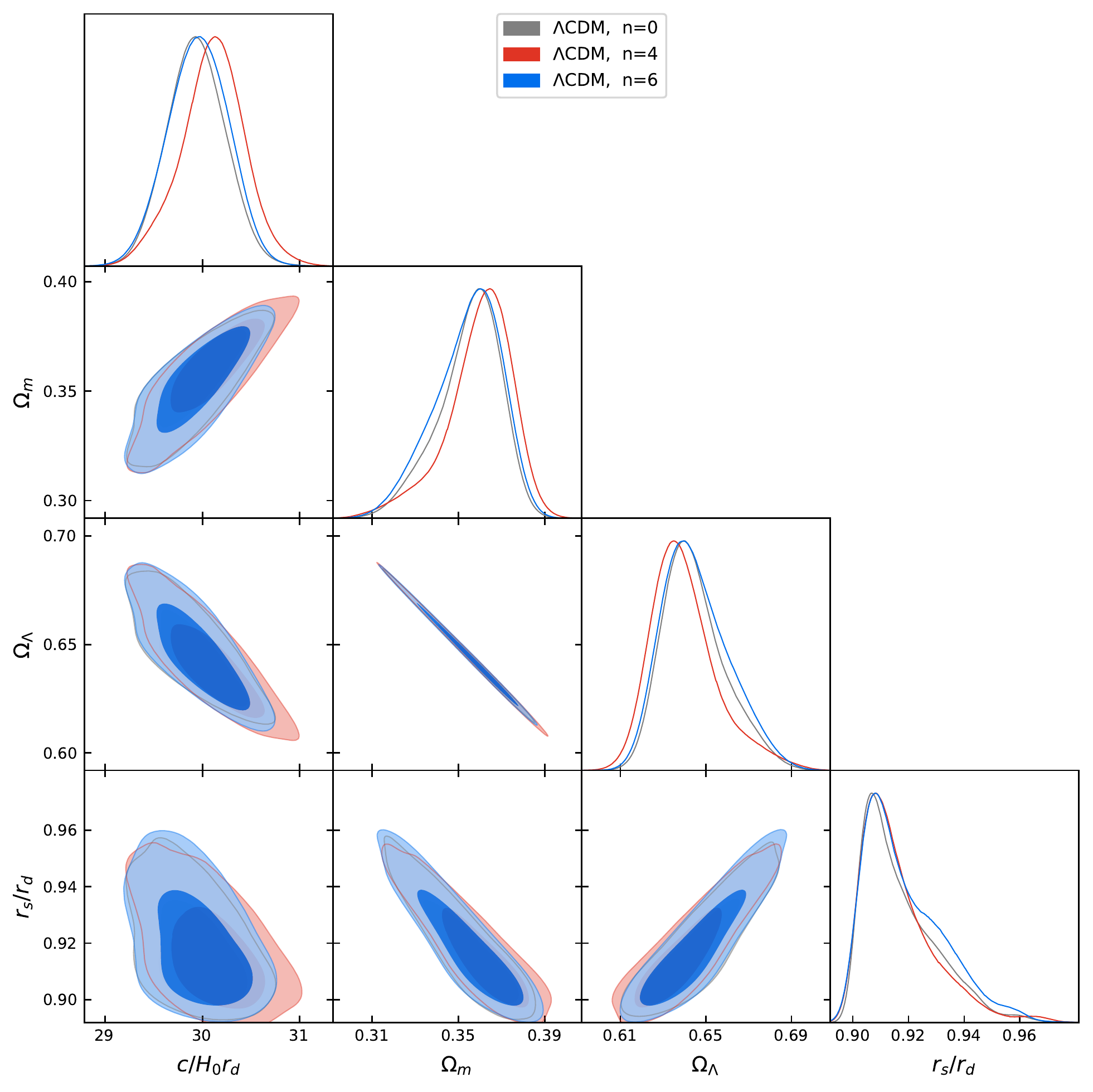}
(\textbf{b}) \includegraphics[width=0.25\textwidth]{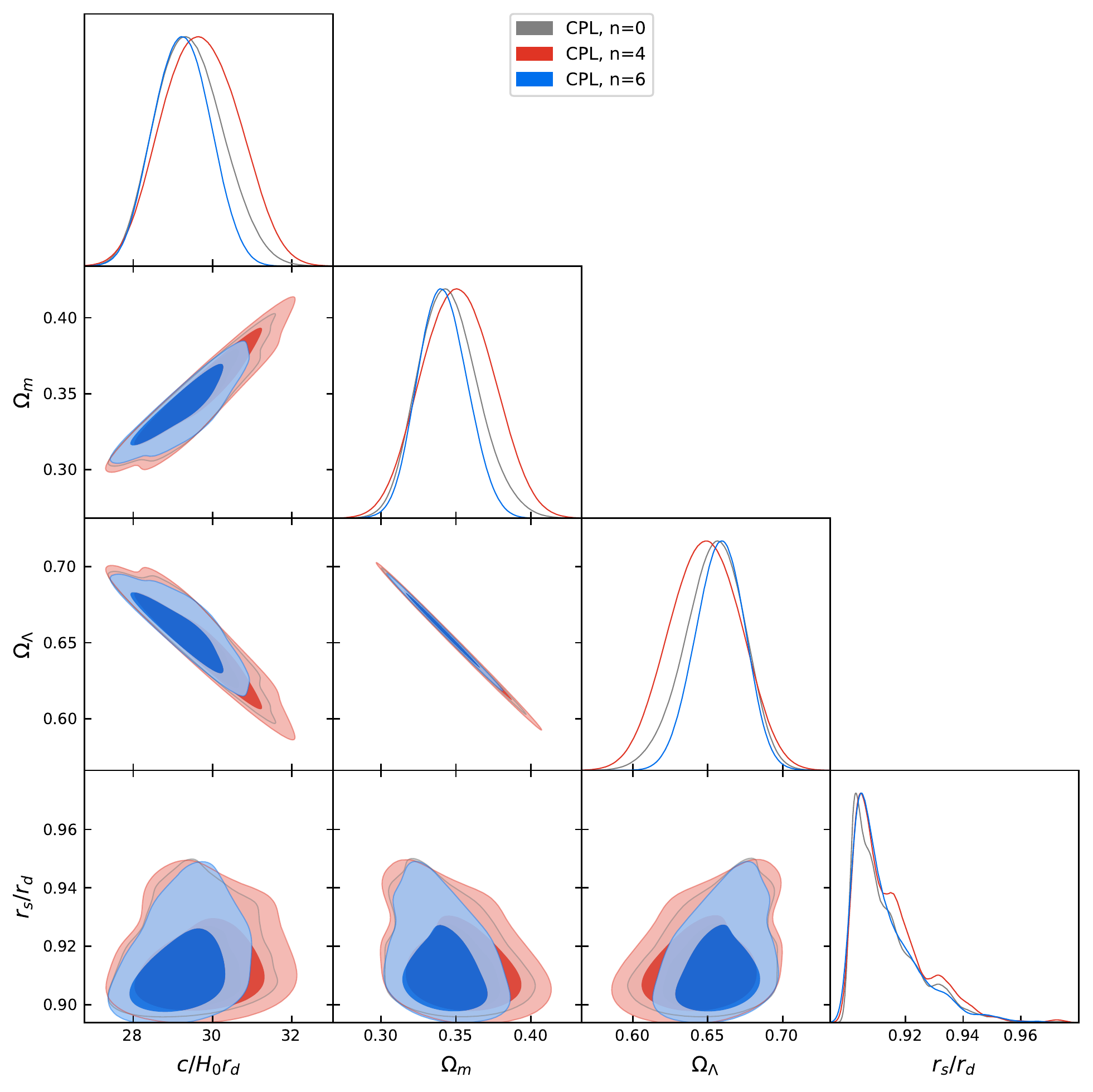}
(\textbf{c}) \includegraphics[width=0.25\textwidth]{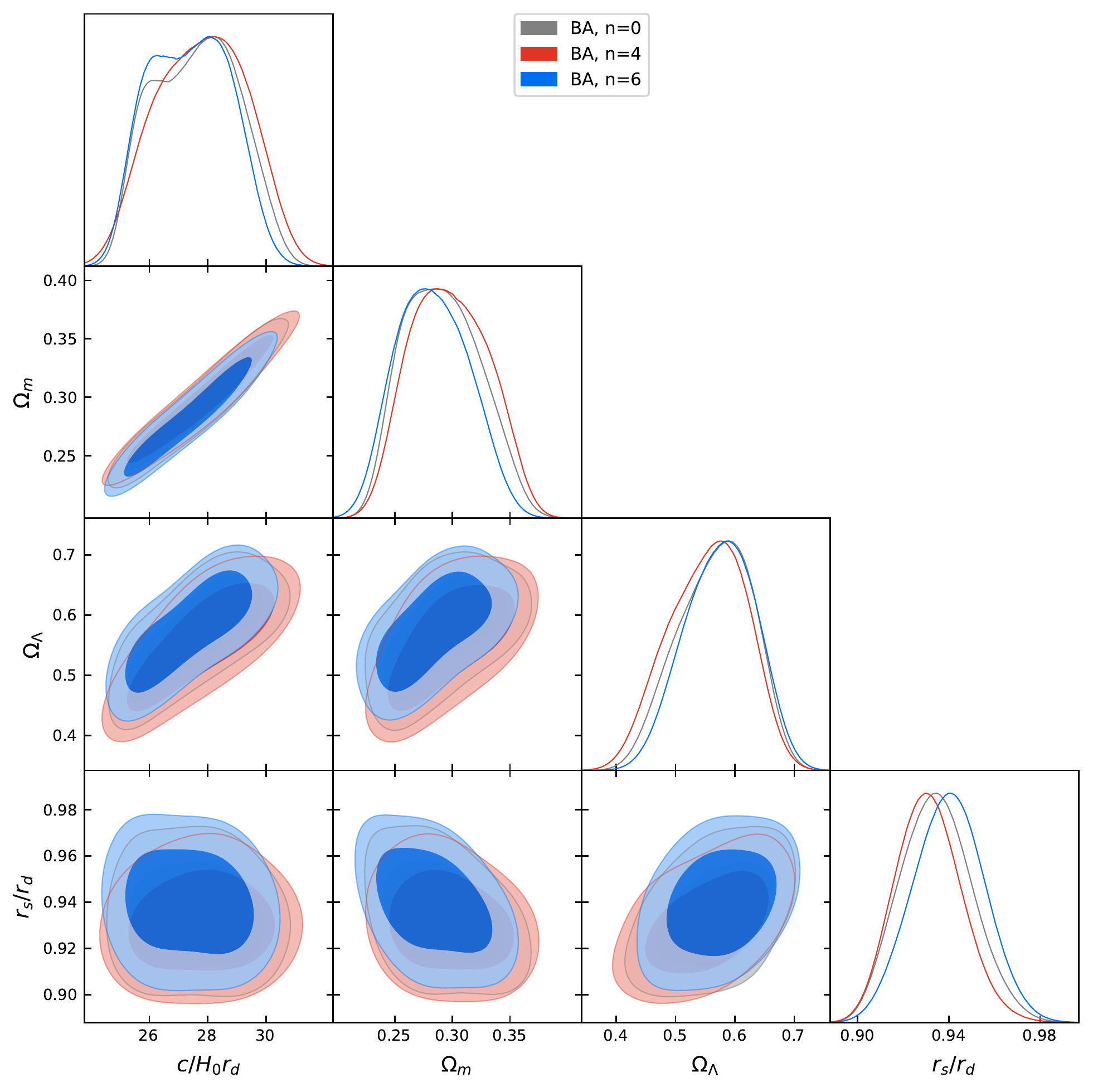}\\
(\textbf{d}) \includegraphics[width=0.25\textwidth]{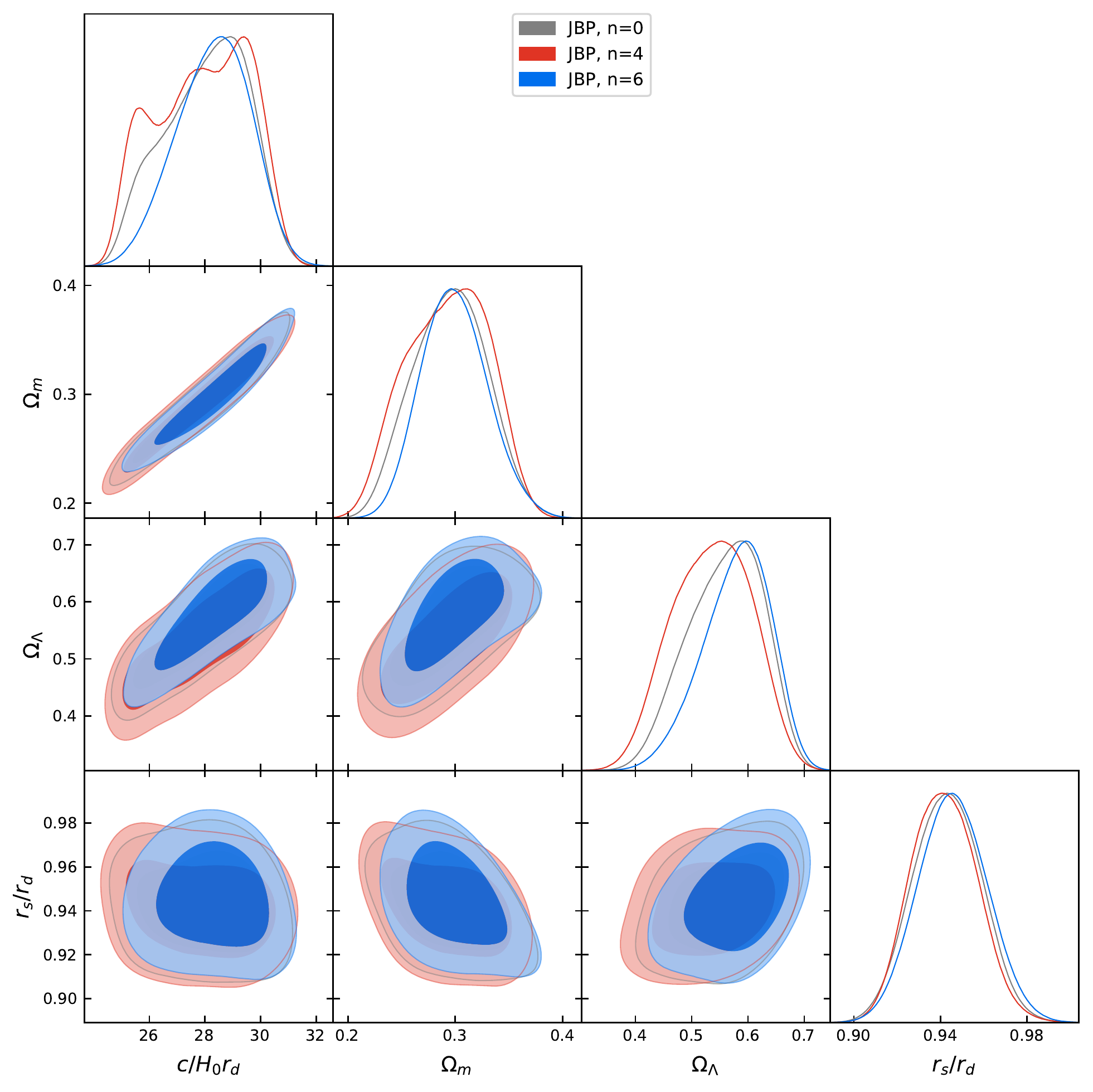}
(\textbf{e}) \includegraphics[width=0.25\textwidth]{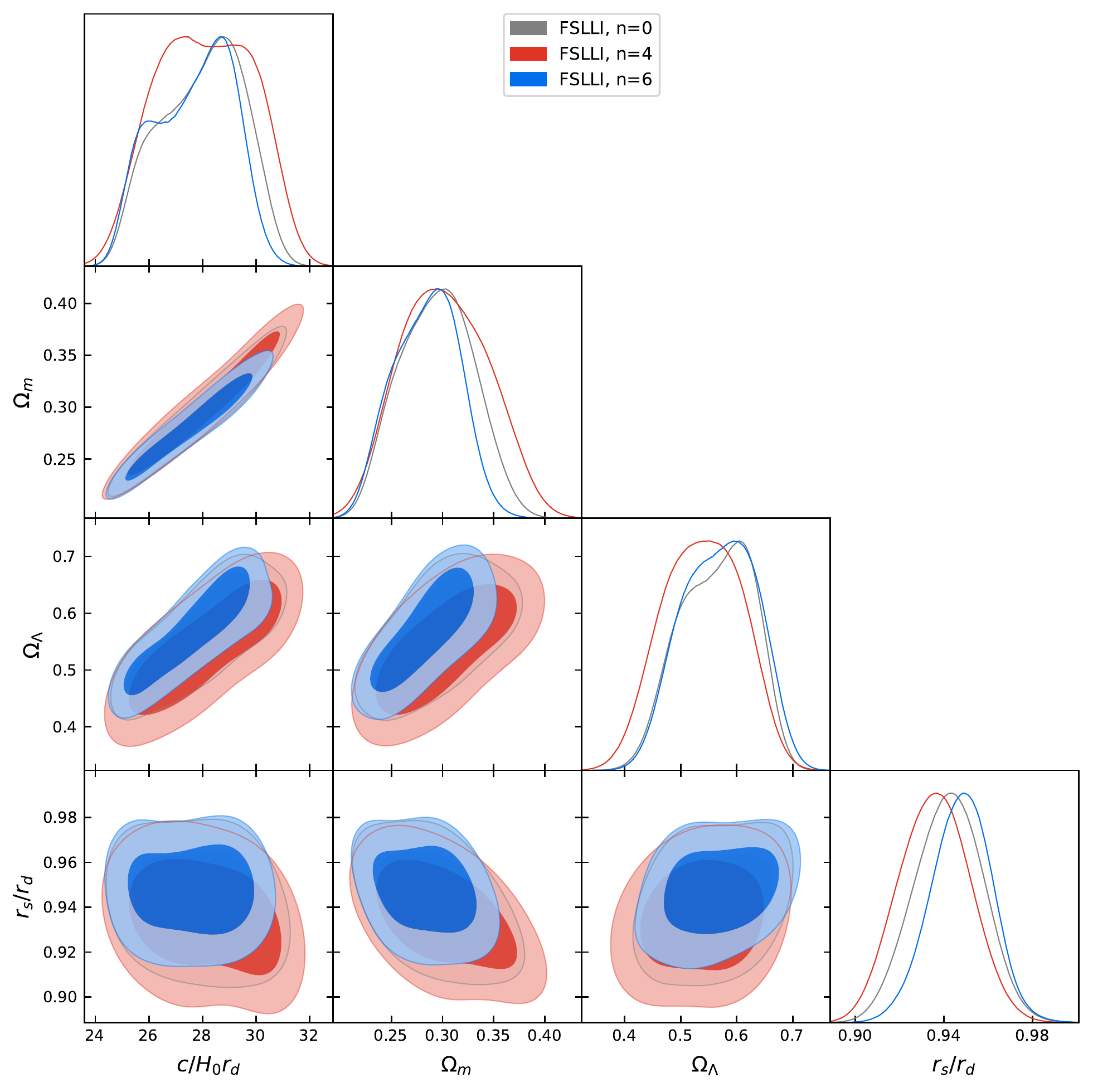}
 \caption{{{The} covariance test plot for the considered models: (\textbf{a}) $\Lambda$CDM, (\textbf{b}) $CPL$, (\textbf{c}) $BA$, (\textbf{d}) $JPB$, (\textbf{e}) $FSLLI$ model.}}
\label{fig:checkCov}
\end{figure}
\unskip

\begin{figure}[H]
 	
\includegraphics[width=0.45\textwidth]{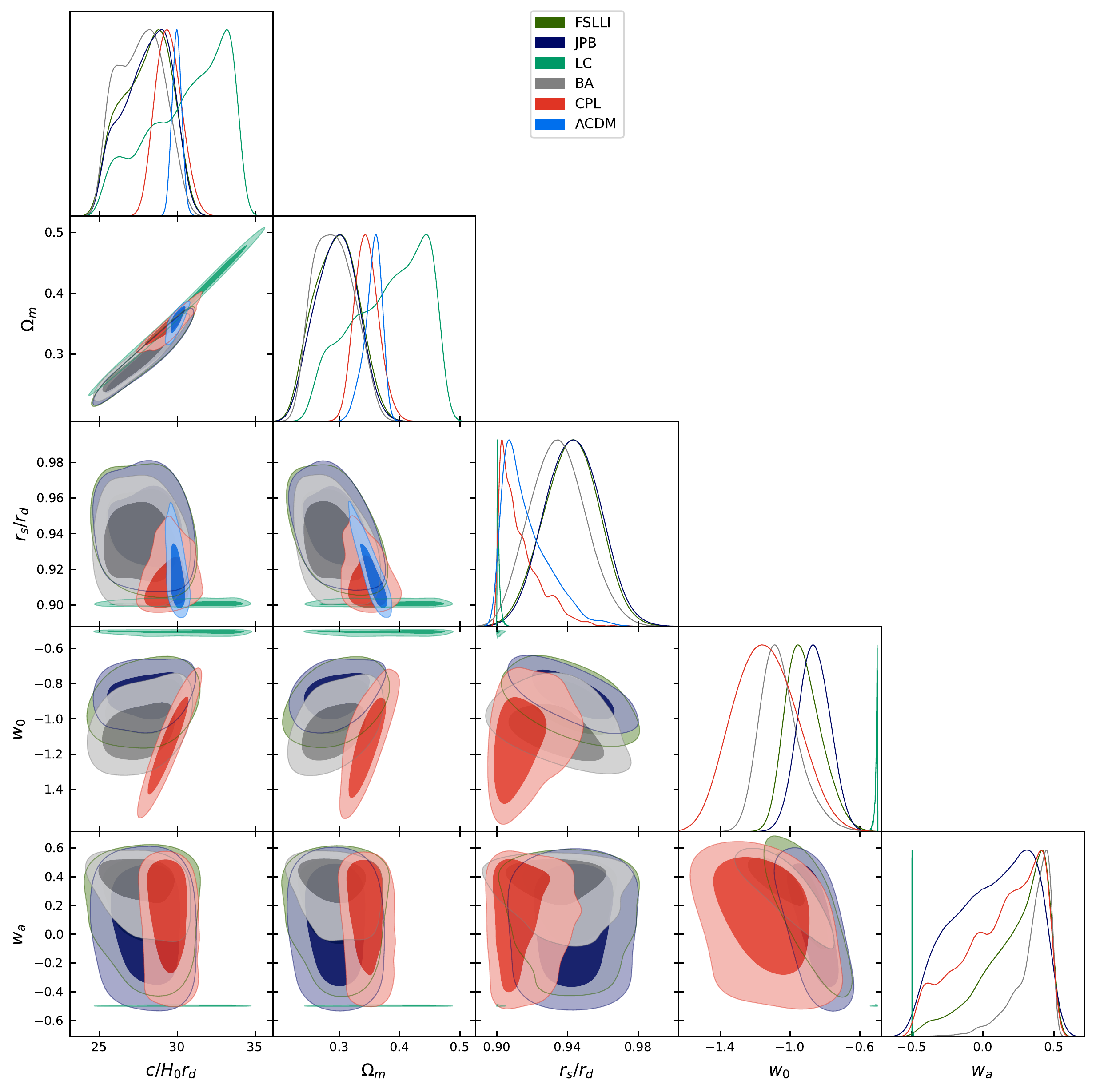}
\includegraphics[width=0.45\textwidth]{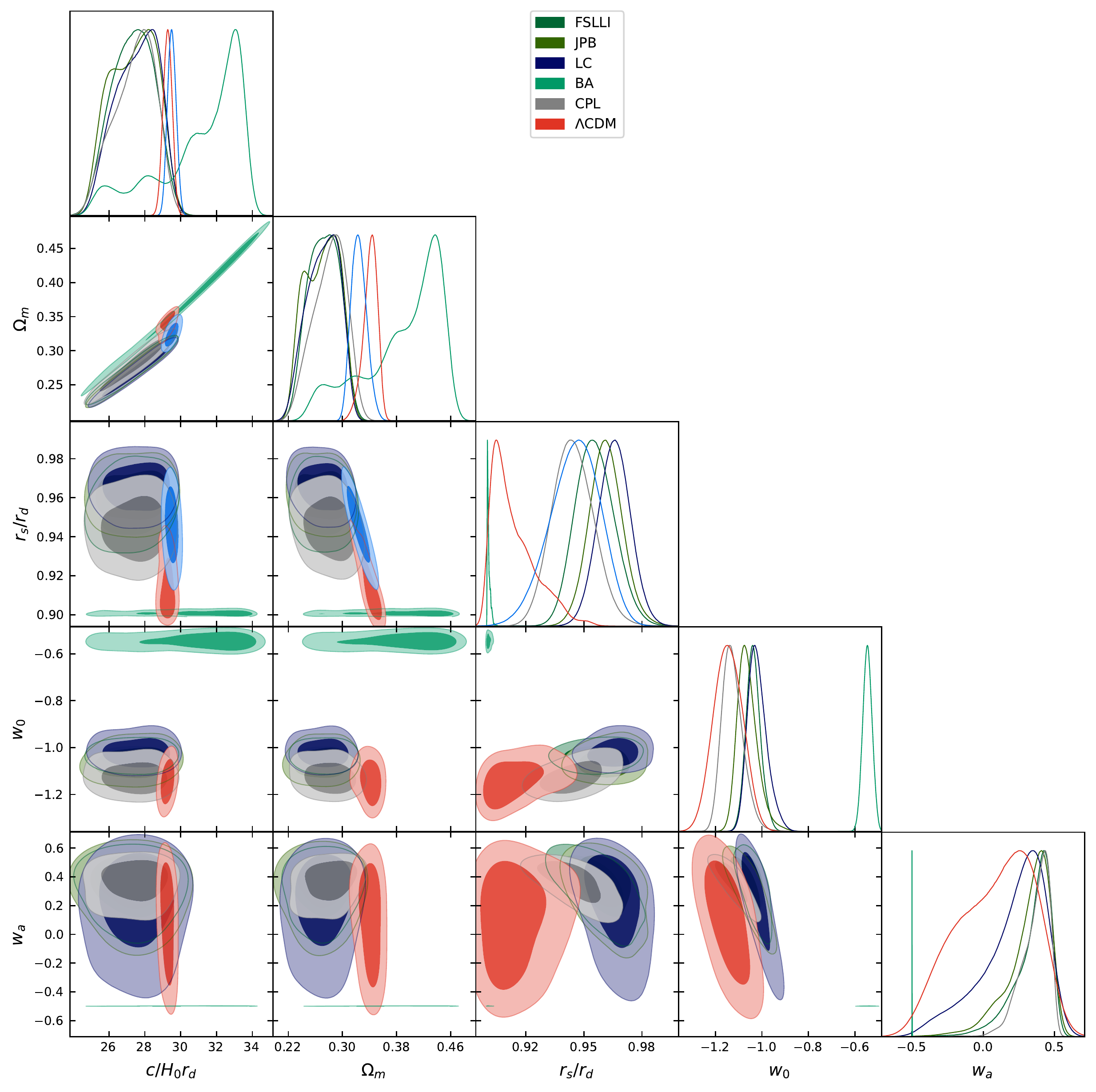}
	\caption{{The} posterior distribution for $c/(H_0 r_d)$, $\Omega_m$, $r_*/r_d$ and $w_0,w_a$ for different parametrization of DE for the BAO+CMB dataset to the left and for the BAO+CMB+SN+GRB to the~right. }
\label{results_BAO}
\end{figure}

\begin{adjustwidth}{-\extralength}{0cm}
\printendnotes[custom]

\reftitle{References}




\end{adjustwidth}
\end{document}